\documentstyle{article}
\title{Why Loops Don't Matter 
\bf   
}
\author{ {\it D.A. Johnston}\\
         Dept. of Mathematics\\
         Heriot-Watt University\\
         Riccarton\\
         Edinburgh, EH14 4AS, Scotland\\ \\
and\\ \\
         {\it P. Plech\' a\v{c}}\\
         Mathematical Institute\\
         24-29 St Giles'\\
         Oxford\\
         OX1 3LB}
\date {15th May 1997}         
\begin{document}
  \maketitle
                      {\Large
                      \begin{abstract}
%
In recent work \cite{1} we have found identical behaviour for various
spin models on ``thin'' random graphs - Feynman diagrams - and the corresponding
Bethe lattices. In this paper we observe that 
the ratios of the saddle point equations in the random graph
approach are identical to the fixed point(s)
of the recursion relations which are used
to solve the models on the Bethe lattice. The loops in the random graphs
thus have no influence on the thermodynamic limit for
such ferromagnetic spin models. 

We consider the 
correspondence explicitly for Ising and
$q$ state Potts models and also note that multi-spin interaction models
on cacti admit a similar correspondence with 
a randomised version of the cacti graphs which contain loops. 
%
                        \end{abstract} }
%
  \thispagestyle{empty}
%
%
  \newpage
%
                  \pagenumbering{arabic}

\section{Introduction}

In a series of papers \cite{1} we have found identical behaviour for various
spin models on ``thin'' random graphs - Feynman diagrams -and the corresponding
Bethe lattices. The key observation for the approach of \cite{1} was made
in \cite{2}, where it was noted that spin models on random graphs
could be treated as the $N \rightarrow 1$ limit of the 
appropriate hermitean matrix model. The $N \rightarrow \infty$ limit
in such models is already familiar from the theory of
$2D$ gravity, where it generates planar random graphs
and the expansion in $1 / N$ is an expansion in the topology of the
graphs. The $N \rightarrow 1$ limit weights all topologies
equally so the graph/surface correspondence of the $N \rightarrow
\infty$ limit, which is important for $2D$ gravity and string theory
applications, is lost. Nonetheless, the generic random graphs
of \cite{1} (which we denoted as ``thin'' graphs to distinguish
them from the ``fat'' graphs of the planar limit) are
still of interest in a statistical mechanical
context because their locally tree-like structure \cite{3} means that one
obtains mean field transitions for spin models which live on them.
They offer one great advantage over genuine tree-like structures such as the 
Bethe \cite{4} lattice in that they have no boundary, so one is not forced
to discard the majority of vertices in a simulation or be subject to
the gymnastics of considering only points ``deep within'' the lattice
analytically.

One remarkable fact that has emerged from all of the simulations and calculations
in \cite{1} is that the thermodynamic behaviour of spin models with
ferromagnetic couplings on the
Feynman diagrams was {\it identical} to that on the corresponding Bethe lattice,
even down to non-universal features such as the critical temperature.
The loops did play a role when one considered antiferromagnetic couplings,
which offered the possibility of frustration and the presence of spin glass
rather than antiferromagnetic ordering at low temperatures. The
similarity was all the more interesting because the methods of solution
for the two classes of model are, at first sight, completely
different. In the random graph models one looks at a saddle point
equation for the ``fields'' of the model and the saddle point action,
which is exact in the $n = \sharp vertices \rightarrow \infty$ limit,
gives the free energy. All the thermodynamic quantities such as the
magnetisation, energy, specific heat and susceptibilities may then be calculated
from this. For the Bethe lattice models on the other hand, the hierarchical structure
of the lattice allows one to construct  a recursion relation for partition
functions on sub-trees that is iterated shell by shell to get the behaviour at
the central point (i.e. ``deep within'' the lattice). The asymptotic behaviour
of this iteration (fixed points, limit cycle, chaotic \ldots) then indicates
the phase structure of the model \cite{5,6} and provides a fruitful tie-in
with the theory of dynamical systems, particularly for multi-spin interaction
and frustrated models.

In this paper we show that the previous heuristically observed 
identity between the Bethe lattice and its random graph equivalent
is explained by the fact that the ratio of the saddle point equations
in the random graph model gives exactly the 
fixed point(s) of the recursion relation in the
Bethe lattice models. The statement that we have made, and repeated,
in various of \cite{1} that the loops are irrelevant 
for ferromagnetic phase transitions on
random graphs is thus put on a firmer footing. In essence,
the content of the saddle point equations for the random graph models
is identical to the recursion relations
of the Bethe lattice, which of course contains no loops by definition.
We demonstrate this explicitly for the Ising and Potts models
and, in order to show that the result is not confined to nearest
neighbour interactions and the Bethe lattice proper, we also
consider a three-site interaction model  \cite{5,5a,7,7a}
on a cactus graph and its randomised equivalent and find exactly
the same equivalence.
For simplicity, we shall consider 3-regular ($\phi^3$) random graphs 
and cacti built from triangles
throughout this paper, which correspond to
a Bethe lattice with 3 neighbours for each site, but the results 
we discuss hold for any number of neighbours.

\section{The Ising Model}

The partition function for the Ising model
with Hamiltonian
\begin{equation}
H = \beta \sum_{<ij>} \sigma_i \sigma_j,
\end{equation}
($\sigma_{i,j} = \pm 1$) on 3-regular random graphs 
\footnote{In principle the thin graph partition functions
are for an annealed ensemble of all graphs with $2n$ vertices.
In practice self-averaging appears to ensure that one can 
consider a {\it single} random graph, at least for ferromagnetic
spin models such as those discussed here.}
with $2n$ vertices may be written as
\cite{1,2}
\begin{equation}
Z_n(\beta) \times N_n = {1 \over 2 \pi i} \oint { d \lambda \over
\lambda^{2n + 1}} \int { {d \phi_+ d \phi_- \over 2 \pi \sqrt{\det K}}
\exp (- A )},
\label{part}
\end{equation}
where $N_n$ is the number of undecorated (no spin) $\phi^3$ graphs
with $2n$ vertices
\begin{equation}
N_n = \left( {1 \over 6} \right)^{2n} { ( 6 n - 1 ) !! \over ( 2 n ) !!
}.
\end{equation}
the action is
\begin{equation}
A = {1 \over 2 } \sum_{a,b}  \phi_a  K^{-1}_{ab} \phi_b  -
{\lambda \over 3} (\phi_+^3 + \phi_-^3).
\label{e3}
\end{equation}
where $a,b = \pm$ and the propagator is
\begin{equation}
\begin{array}{cc} K_{ab} = & \left(\begin{array}{cc}
\sqrt{g} & { 1 \over \sqrt{g}} \\
{1 \over \sqrt{g}} & \sqrt{g}
\end{array} \right) \end{array}
\end{equation}
where $g = \exp (2 \beta)$ relates the coupling in the action
to the inverse temperature $\beta$.

The saddle point equations $\partial A / \partial \phi_{\pm} = 0$
may be written, after scaling out $\lambda$, as
\begin{eqnarray}
\phi_+ &=&  \sqrt{g} \phi_+^2 + {1 \over \sqrt{g}} \phi_-^2 \nonumber \\
\phi_- &=&  \sqrt{g} \phi_-^2 + {1 \over \sqrt{g}} \phi_+^2
\label{saddle}
\end{eqnarray}
and one finds the solutions
\begin{eqnarray}
\phi_+,\phi_- &=&  { ( g - 1 )  \over
         g}  \nonumber \\
\phi_+,\phi_- &=& { 1 + g \pm \sqrt{ (g + 1) ( g  - 3 ) }.
\over 2   g  }
\label{saddlesol}
\end{eqnarray}
The upper high temperature paramagnetic solution bifurcates
to the low temperature ferromagnetic solution
in a mean-field transition at $g=3$, as can be seen by looking at
the magnetisation order parameter, which in this notation is
\begin{equation}
M = { \phi_+^3 - \phi_-^3 \over \phi_+^3 + \phi_-^3}.
\end{equation}

At first sight these solutions, and the method by which they were 
obtained bear little relation to the Bethe
lattice calculation \cite{5} where
the partition function may be expressed in a recursive
manner as
\begin{equation}
Z = \sum_{\sigma_0} \exp ( h \sigma_0 ) \left[ g_n (\sigma_0) \right]^3
\end{equation}
where the central spin is $\sigma_0$, the external field is $h$,
we have $n$ ``shells'' in the Bethe lattice and 
\begin{equation}
g_n (\sigma_0) = \sum_s Q_n ( \sigma_0 \vert s )
\end{equation}
with
\begin{equation}
Q_n ( \sigma_0 \vert s ) = \exp ( \beta \sigma_0 s_1 + h s_1 )
\prod_{j=1}^{2} Q_{n-1} ( s_1 \vert t^{(j)} )
\end{equation}
reflecting the decomposition of the tree into trunk and branches.
$s$ denotes all the spins
apart from $\sigma_0$ on the $j$th sub-tree and 
$t^{(j)}$ denotes the spins apart from $s_1$ on the $j$th branch
of the sub-tree.
Letting
\begin{equation}
x_n = { g_n ( - ) \over g_n ( + )},
\end{equation}
and setting the external field $h$ to zero, we find
finally
\begin{equation}
x_{n+1} = { \frac{1}{\sqrt{g}} + \sqrt{g} x_n^2 \over \sqrt{g} + \frac{1}{\sqrt{g}} x_n^2}.
\label{iter}
\end{equation}
This recursion relation gives a single fixed point in the paramagnetic phase ($g<3$)
and a pair of stable fixed points with one unstable fixed point in the ferromagnetic phase
($g>3$).

Although the phase transition points are identical they have apparently appeared
by very different paths on the Bethe lattice and on the random graphs. However,
a much closer parallel appears between equ.(\ref{saddle}) and equ.(\ref{iter})
when one divides the saddle point equation for $\phi_-$ by that for $\phi_{+}$ to get
\begin{equation}
{\phi_- \over \phi_+} =  { \sqrt{g} \phi_-^2  + \frac{1}{\sqrt{g}} {\phi_+^2} 
                          \over \sqrt{g}  \phi_+^2  + \frac{1}{\sqrt{g}} \phi_-^2} 
\end{equation}
which is identical to  equ.(\ref{iter}) when we set $x_n = \phi_- / \phi_+$ and go to the fixed point(s).
The loops in the random graphs have thus made no contribution to the 
saddle point equations, which are completely equivalent to the fixed point solutions
of equ.(\ref{iter}). 

Although we have only detailed here the results when the external field is zero,
the divided saddle point equations are still identical to the Bethe lattice recursion
relation at the fixed point in non-zero field when the identification
$x = \phi_- / \phi_+$ is made. As a consequence, quantities such as the magnetisation
are also given by identical formulae on thin graphs and on the Bethe lattice
(for a site ``deep within'' the lattice). 
The Ising model is the simplest decoration that we could consider for our
Bethe lattice and $\phi^3$ random graphs, but the correspondence
between the critical behaviour on the two classes of lattices is no fluke and is maintained
for other spin models, as we see in the next two sections.

\section{The Potts Model}

The $q$ state Potts model Hamiltonian may be taken to be
\begin{equation}
H =   \beta \sum_{<ij>} ( \delta_{\sigma_i, \sigma_j} -1),
\end{equation}
where the spins $\sigma_i$ now take on $q$ different values.
The general formula for the partition function is similar to the Ising
model, but there are now $q$ fields $\phi$ and the action is given by
\begin{equation}
A = { 1 \over 2 } \sum_{i=1}^{q} \phi_i^2 - c \sum_{i<j} \phi_i \phi_j -{\lambda \over 3} \sum_{i=1}^q \phi_i^3,
\label{qstate}
\end{equation}
where the coupling is related to the inverse temperature by
\begin{equation}
c = { 1\over  ( \exp( 2 \beta)  + q-2) }.
\label{coup}
\end{equation}
One finds the low temperature saddle point solutions
\begin{eqnarray}
\phi_{1 \ldots q-1} &=& { 1 -  (q-3) c - \sqrt{1 - 2 (q-1) c + (q-5) (q-1) c^2} \over 2}
\nonumber \\ 
\phi_q &=& { 1 +  (q-1) c + \sqrt{ 1 - 2 (q-1) c + (q-5) (q-1) c^2} \over 2}
\label{qsols}
\end{eqnarray}
(and another low temperature branch with the signs in front of the square roots reversed)
which can be obtained self-consistently by imposing the observed symmetry breaking pattern
on the full action in equ.(\ref{qstate}) to get
\begin{eqnarray}
A = {1 \over 2} ( q - 1) \left[ 1 - c ( q - 2) \right] \phi^2  - { 1 \over 3} ( q -1) \phi^3
+ {1 \over 2} \tilde \phi^2  - {1 \over 3} \tilde \phi^3 - c ( q -1 ) \phi \tilde \phi
\label{app1}
\end{eqnarray}
with $\phi=\phi_{1 \ldots q-1}$, $\tilde \phi = \phi_q$
and $\lambda$ scaled out.
Similarly, 
at high temperature
the saddle point solution is $\phi_{1 \ldots q} = \phi_0 = 1 - ( q - 1) c$.
which can be obtained from the effective action
\begin{eqnarray}
A_0 = { q \over 2} (1 - c ( q - 1) ) \phi_0^2 - { q \over 3} \phi_0^3.
\label{app2}
\end{eqnarray}
Although we started with $q$ fields the final saddle point solution
only requires two fields in its solution
as a consequence of the symmetry breaking pattern in the low temperature phase,
just as for the Ising model.

The ratio of the saddle point equations following from equ.(\ref{app1})
may be written as
\begin{eqnarray}
{\phi \over \tilde \phi} = { \phi^2  + c \tilde \phi^2 \over c ( q -1 ) \phi^2 + ( 1 - c ( q - 2) ) \tilde \phi^2 }  
\label{saddlep}
\end{eqnarray}
and it is not immediately apparent that this is equal to the recursion relations for
the Potts model on a Bethe lattice in \cite{8}, which may be written in the form 
\begin{eqnarray}
x_{n+1} = \left[ { 1 + x_n / c \over \exp(2 \beta ) + (q - 1) x_n } \right]^2.
\label{curp}
\end{eqnarray}
However, one can rearrange equ.(\ref{saddlep})
into
\begin{eqnarray}
z = { 1 + z^2 / c \over ( q - 1) z^2 + \exp(2 \beta ) }
\end{eqnarray}
where $z = \phi / \tilde \phi$.
If we now make the identification $z^2 = x$ we recover the fixed point of the
recursion relation in equ.(\ref{curp}).
We have thus found, just as for the Ising model, (which is, after all, a $q=2$ state
Potts model) that the fixed points of the Bethe lattice recursion relations
and the saddle point equations on thin graphs are identical. 

\section{Multi-site interaction models and
Loopy Cacti}

A Husimi Tree (or ``cactus'' graph) is a tree-like structure
in the large rather like the Bethe lattice, but the individual
components are polygons joined at their vertices, rather
than simply vertices and links. It is characterised
by the number of polygons articulated at each vertex, $\gamma$.
An cactus with $\gamma=2$
is shown in Fig.1 along with the 
Bethe lattice for which it forms a sort of medial graph.
For $\gamma>2$ one does not have such a direct
correspondence but the hierarchical structure still, of course, remains
which allows
one to deploy similar techniques to the 
Bethe lattice solution, using
recursion relations that operate shell by shell.
Since the individual units are now more complicated than
vertices, it is possible to put models with
multi-site interactions on such trees and investigate their phase
behaviour \cite{6,7,7a}. Writing down a thin graph style action
for a such models corresponds to allowing (
predominantly large) loops
in the branches of the cacti, in the same manner as the Feynman diagram
expansion introduces large loops into a locally tree like
structure. Given the results of the previous sections 
relating the critical behaviour on the Bethe lattice and thin graphs one
might expect that the loopy cacti would display
exactly the same behaviour as the loop-less cacti.

As an example we take the following multi-spin interaction 
model (a special case of the model solved by Wu and Wu \cite{5a}
on the Kagom\'e lattice) 
\begin{equation}
H = \beta ( J_3 \sum_{\Delta} \sigma_i \sigma_j \sigma_k + h \sum_i \sigma_i)
\end{equation}
where the three spin sum is over triangles $\Delta$.
If one considers the possible spin configurations on the triangle
of the Husimi tree, namely
$(+ \; + \; + ), \; (-\; -\; -), \; (+ \; + \; -), \; (+ \; - \; -)$,
the appropriate action can be written for the
lattice which looks locally like Fig.1 
(but has large loops of triangles) by thinking of the spins
as residing on the links of the underlying $\phi^3$ graph. One then
has ``ghost'' links joining ``ghost'' vertices at the centre
of each of the triangles, which gives
\begin{equation}
A = {\mu \over 2 }  \phi_+^2 + {1 \over 2} \phi_-^2 -  \mu^2  \phi_+^2 \phi_- 
    - z \mu  \phi_-^2 \phi_+ - { z \mu^3 \over 3} \phi_+^3 - {1 \over 3} \phi_-^3,
\end{equation}
where $\mu = \exp ( 2 \beta h)$, $z = \exp ( 2 \beta J_3)$
and we have again 
pre-emptively scaled out the vertex counting factor $\lambda$
for simplicity.
Note that an edge on which
a positive spin resides, represented by the ``propagator'' for $\phi_+$, picks up a factor of $1 / \mu$
to avoid double counting the external field contribution 
where two triangles join.
If we now consider the saddle point equations for the model
\begin{eqnarray}
\phi_+ &=&  z \mu^2 \phi_+^2 + 2 \mu \phi_+ \phi_- + z \phi_-^2 \nonumber \\
\phi_- &=&  \mu^2 \phi_+^2 + 2 z \mu \phi_+ \phi_- +\phi_-^2 
\label{saddlecac}
\end{eqnarray}
we find that their ratio is again exactly the recursion relation
derived for the model on a loop-less cactus (Husimi tree) in \cite{7a}.

The Hamiltonian may be extended to include nearest neighbour interactions as well
\begin{equation}
H = \beta ( J_3 \sum_{\Delta} \sigma_i \sigma_j \sigma_k + J_2 \sum_{ij} \sigma_i \sigma_j + h \sum_i \sigma_i).
\end{equation}
and similar reasoning to the pure 3-spin interaction case above leads to the action
\begin{equation} 
A = \nu \left( {\mu \over 2 }  \phi_+^2 + {1 \over 2} \phi_-^2 -  \mu^2 
 \phi_+^2 \phi_-
    - z \mu  \phi_-^2 \phi_+ - { z \nu^2 \mu^3 \over 3} \phi_+^3 - {\nu^2  \over 3} \phi_-^3 \right),
\end{equation}
for the model on loopy cacti,
where the additional parameter $\nu=\exp ( 2 \beta J_2)$.
The overall factor of $\nu$ is irrelevant for solving the saddle point equations
and we find that their ratio is identical to the recursion relation presented
for the model on a loop-less cactus in \cite{7}
\begin{eqnarray}
{\phi_+ \over \phi_-} =  { z \mu^2 \nu^2 \phi_+^2 + 2 \mu \phi_+ \phi_- + z \phi_-^2 \over
            \mu^2 \phi_+^2 + 2 z \mu \phi_+ \phi_- + \nu^2 \phi_-^2}
\end{eqnarray}
The loops in the cacti have thus not altered 
the critical behaviour of the models from
the loop-less case, at least for ferromagnetic couplings, where one is looking for fixed points
of the recursion relations rather than more exotic behaviour.

\section{Discussion}

We have seen that for all the models considered - Ising, Potts,
multi-site interaction, and mixed interaction, the
randomisation of the Bethe lattice or Husimi tree
by the introduction of loops
has no effect on the thermodynamic limit for ferromagnetic
phase transitions. 
This irrelevance of loops is also to be expected on graph
theoretic grounds \cite{3}, for it is known that the loop
distribution on a regular random graph is very strongly weighted 
to large loops \footnote{More precisely, 
on 3 regular graphs the number of
$i$-cycles, that is loops with $i$ edges, is asymptotically
a Poisson random variable with mean $2^i/ (2i)$.}.
This provides a justification for calling the thin graphs
locally tree-like.

The dynamical systems aspects of the
recursive approach on
Bethe lattices and Husimi trees are lost in the ratios 
of the saddle point equations when one 
moves to randomised structures, so 
phases with staggered order
(in particular, antiferromagnetic order) which appear as cycles
in the solution of the recursion relations are  also
absent.
This is intuitively reasonable, because the $\phi^3$ graphs
contain both odd and even loops, so the shell by shell alternation
of spin values that characterises antiferromagnetic order
on the Bethe lattice proper would lead to frustration via the loops.

Indeed, there is strong evidence \cite{1,2} both analytical and numerical that
the Ising model with purely antiferromagnetic couplings displays
a spin glass phase at sufficiently low temperature on $\phi^3$ graphs
rather than the antiferromagnetic phase of the Bethe lattice.
It would be extremely interesting to investigate some of the
more esoteric dynamical phenomena in the recursion relations
of the multi-site interaction models with antiferromagnetic
couplings that are highlighted
in \cite{7,7a} in order to see what, if any, phenomena they
corresponded to in the randomised/loopy case.

\bigskip

\clearpage \newpage
\begin{figure}[htb]
\vskip 20.0truecm
\includegraphics{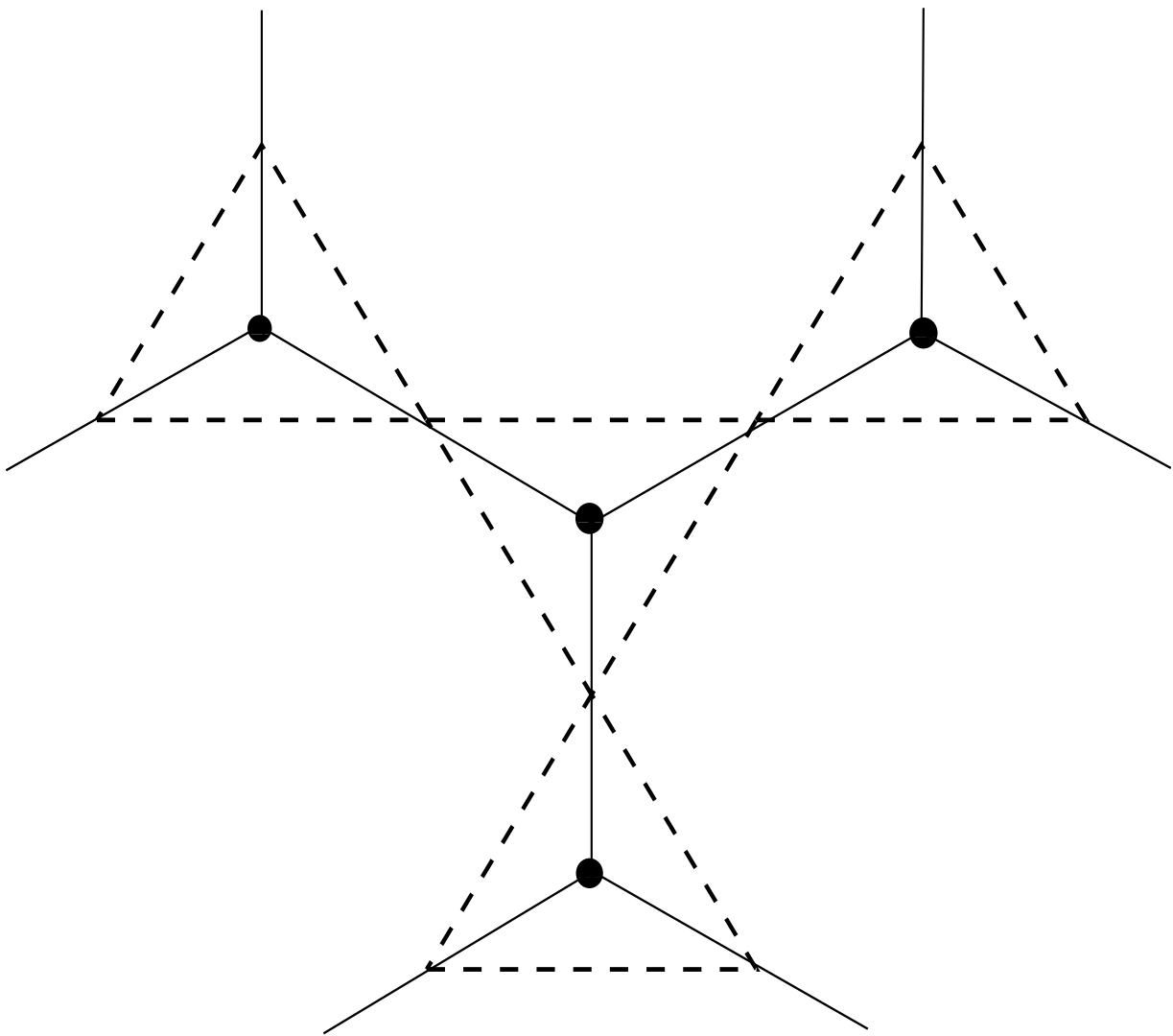}
\caption[]{\label{fig1} A section of a Husimi tree (dotted)
composed of triangles with $\gamma=2$. The underlying
Bethe lattice (solid) is also shown, along with the ``ghost'' nodes at
the centre of each triangle.}
\end{figure}


\begin{thebibliography}{99}
\bibitem{1} D. Johnston and P. Plech\' a\v{c}, ``Potts Models on Feynman Diagrams'',
[hep-lat/9704020];\\
D. Johnston and P. Plech\' a\v{c}, ``Percolation on a Feynman Diagram'',
[cond-mat/9705101];\\
            C. Baillie, D. Johnston and J-P. Kownacki, Nucl. Phys. {\bf B432} (1994) 551;\\
            C. Baillie, W. Janke, D. Johnston and P. Plech\' a\v{c}, Nucl. Phys. {\bf B450}
(1995) 730;\\
            C. Baillie and D. Johnston, Nucl. Phys. {\bf B47} (Proc. Suppl.) (1996) 649;\\
            C. Baillie, D. Johnston, E. Marinari and C. Naitza, J. Phys. {\bf A29} (1996) 6683;\\
            C. Baillie, N. Dorey, W. Janke and D. Johnston, Phys. Lett {\bf B369} (1996) 123.
\bibitem{2} C. Bachas, C. de Calan and P. Petropoulos, J. Phys. {\bf A27} 
            (1994) 6121.
\bibitem{3} B. Bollob\'as, ``Random Graphs'', Academic Press, 1985.
\bibitem{4} H. A. Bethe, Proc. Roy. Soc. {\bf A 150} (1935) 552;\\
            C. Domb, Advan. Phys. {\bf 9} (1960) 145;\\
            T. P. Eggarter, Phys. Rev. {\bf B9} (1974) 2989;\\
            E. Muller-Hartmann and J. Zittartz, Phys. Rev. Lett. {\bf
            33} (1974) 893.
\bibitem{5} R. Baxter, ``Exactly Soluble Models in Statistical Mechanics'', Academic Press, London,
1982.
\bibitem{5a} X. Wu and F. Wu, J. Phys. {\bf A 22} (1989) L1031;\\
             R. Baxter and F. Wu, Phys. Rev. Lett {\bf 31} (1973) 1294.
\bibitem{6} C. Thompson, J. Stat. Phys. {\bf 27} (1982) 441; {\it ibid} 457.
\bibitem{7} J. Monroe, J.Stat. Phys. {\bf 65} (1991) 255; {\it ibid} {\bf 67} (1992) 1185;\\
            P Gujrati, Phys. Rev. Lett. {\bf 74} (1995) 809.
\bibitem{7a}N. Ananikian, S. Dallakian, N. Izmailian and K. Oganessyan, Phys. Lett. {\bf A214} (1996) 205;\\
            A. Alahverdian, N. Ananikian, S. Dallakian, ``Singularities at a Dense Set of Temperature
            in Husimi Tree'', cond-mat/9702106.
\bibitem{8} F. Peruggi, J. Phys. {\bf A16} (1983) L713.\\
             F. Peruggi, F. di Liberto and G. Monroy, J. Phys. {\bf A16}
             (1983) 811;\\
             F. Peruggi, Physica {\bf 141A} (1987) 140;\\
             F. Peruggi, F. di Liberto and G. Monroy, Physica {\bf 141A} (1987) 151;\\
               F. Peruggi, F. di Liberto and G. Monroy, Z. Phys. {\bf B66}, (1987) 379;\\
               J. Essam, J-C Lin and P. Taylor, Phys. Rev. {\bf E52} (1995) 44.
          
\end{thebibliography}
\end{document}